\documentclass[a4paper]{article}
\usepackage{amsmath,amssymb,graphicx}
\newcommand{\st}{\hbox{$*$}}
\newtheorem{proposition}{Proposition}
\everymath{\displaystyle}
\begin{document}
\title{Asymptotic solution to a fuzzy elementary cellular automaton of rule number 38}
\author{Ko Yamamoto, Daisuke Takahashi\\
\ \\
Department of Pure and Applied Mathematics, Waseda University,\\
3-4-1, Okubo, Shinjuku-ku, Tokyo, 169-8555, Japan}
\date{}
\maketitle
\begin{abstract}
  Fuzzy cellular automaton is a dynamical system with a continuous state value embedding a cellular automaton with a discrete state value.  We investigate a fuzzy cellular automaton obtained from an elementary cellular automaton of rule number 38.  Its asymptotic solutions are classified into two types.  One is a solution where stable propagating waves exist, and the other is a static uniform solution of constant value.
\end{abstract}
\section{Introduction}
Cellular automaton (CA) is a dynamical system with a finite set of state values in discrete coordinates.  Various theories and applications have been done about CA's due to their concise construction and rich expression capability\cite{wolfram}.  One of the simplest configurations is Elementary CA (ECA) of which binary state value at next time is determined by three neighbors in one-dimensional space sites at the current time as follows:
\begin{equation}  \label{eca}
  u_j^{n+1}=f(u_{j-1}^n,u_j^n,u_{j+1}^n),
\end{equation}
where $j$ denotes an integer space site, $n$ an integer time step and $u$ a binary state value ($u\in\{0,1\}$).  Since $f$ is binary-valued with three binary arguments, it can be defined by the following rule table where $b_k\in\{0,1\}$.
\begin{equation}  \label{rule}
\begin{array}{|c|c|c|c|c|c|c|c|c|}
\hline
x\,y\,z & 111 & 110 & 101 & 100 & 011 & 010 & 001 & 000 \\
\hline
f(x,y,z) & b_1 & b_2 & b_3 & b_4 & b_5 & b_6 & b_7 & b_8 \\
\hline
\end{array}
\end{equation}
Thus, there are 256 different rules defined by the above rule table and every ECA is distinguished by the rule number $(b_1b_2\ldots b_8)_2$.  They have been studied theoretically from various viewpoints; mathematical structure of solutions, statistical mechanism on solution patterns, and so on.\par
  The systems obtained by embedding CA in continuous real or rational background is generally called `fuzzy' cellular automaton\cite{cattaneo}.  For example, fuzzy ECA is defined in the form of (\ref{eca}) where $j$ and $n$ are integer and $u\in[0,1]$.  There are infinite variations on $f$ since its necessary condition is $[0,1]^3\to[0,1]$ together with $\{0,1\}^3\to\{0,1\}$.  This condition means fuzzification together with embedding of CA.  One of popular forms of fuzzy CA is defined by using the polynomial\cite{betel,mingarelli}.  For example, if we define
\begin{equation}
  f(x,y,z)=xyz,
\end{equation}
then (\ref{eca}) becomes fuzzy ECA of rule number 128.\par
  Since fuzzy CA is a continuity extension of CA, it has been used as application models to express an intermediate state value among the original discrete values\cite{liu,bone}.  Moreover, there exist another important significance for fuzzy CA from the theoretical viewpoint.  Since it embeds CA in the continuous range, continuous solutions to fuzzy CA propose a rich comprehension to discrete ones to its original CA\cite{sakata}.  We discuss asymptotic solutions to a fuzzy CA obtained from an ECA in this article.  Its range is a continuous interval $[0,1]$ and it also proposes solutions to the ECA as a special case of a discrete range $\{0,1\}$.\par
\begin{figure}[hbt]
\begin{center}
  \includegraphics[scale=0.5,bb=0 0 360 256]{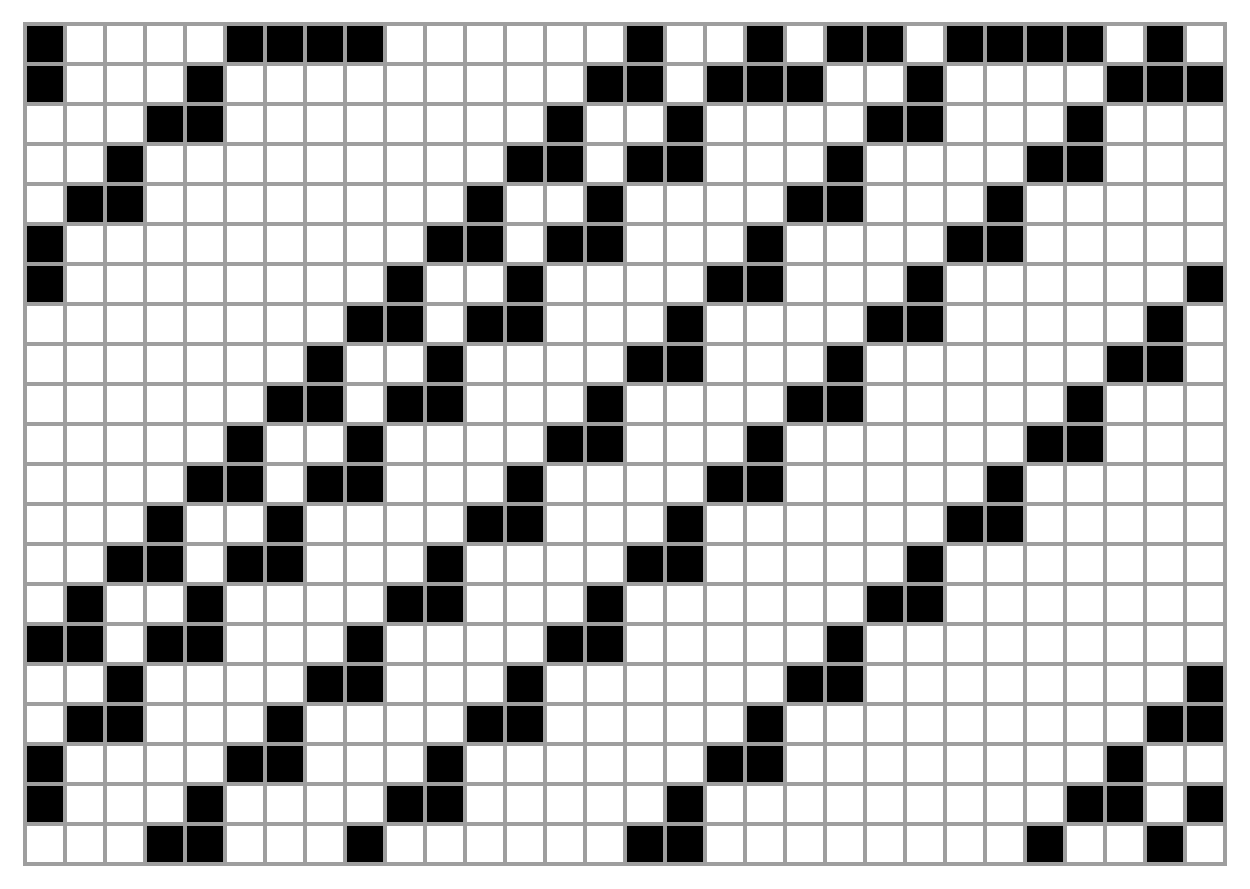}
\end{center}
  \caption{Example of solution to ECA38 for $K=30$.  The space coordinate $j$ and $n$ are rightward and downward respectively.}
  \label{fig:ECA38}
\end{figure}

  Let us consider the following equation.
\begin{equation}  \label{fuzzy ECA}
\begin{aligned}
  u_j^{n+1}&=f(u_{j-1}^n,u_j^n,u_{j+1}^n),\\
f(x,y,z)&=y+z-xy-2yz+xyz\\
&=(1-x)y(1-z)+(1-y)z,
\end{aligned}
\end{equation}
where $j$ denotes integer space site and $n$ integer time step.  Space is finite and the domain is $0\le j<K$ with a periodic boundary condition $u_{j+K}^n=u_j^n$.  We can easily show the value of solutions to (\ref{fuzzy ECA}) can be closed in $u\in\{0,1\}$, $(0,1)$, or $[0,1]$.  If $u\in\{0,1\}$, then (\ref{fuzzy ECA}) is equivalent to the following rule table.
\begin{equation}  \label{ECA38}
\begin{array}{|c|c|c|c|c|c|c|c|c|}
\hline
x\,y\,z & 111 & 110 & 101 & 100 & 011 & 010 & 001 & 000 \\
\hline
f(x,y,z) & 0 & 0 & 1 & 0 & 0 & 1 & 1 & 0 \\
\hline
\end{array}
\end{equation}
It is the evolution rule of ECA of rule number 38 and an example of time evolution is shown in Figure~\ref{fig:ECA38}.
\par
\begin{figure}[hbt]
\begin{center}
  \includegraphics[scale=0.5,bb=0 0 360 256]{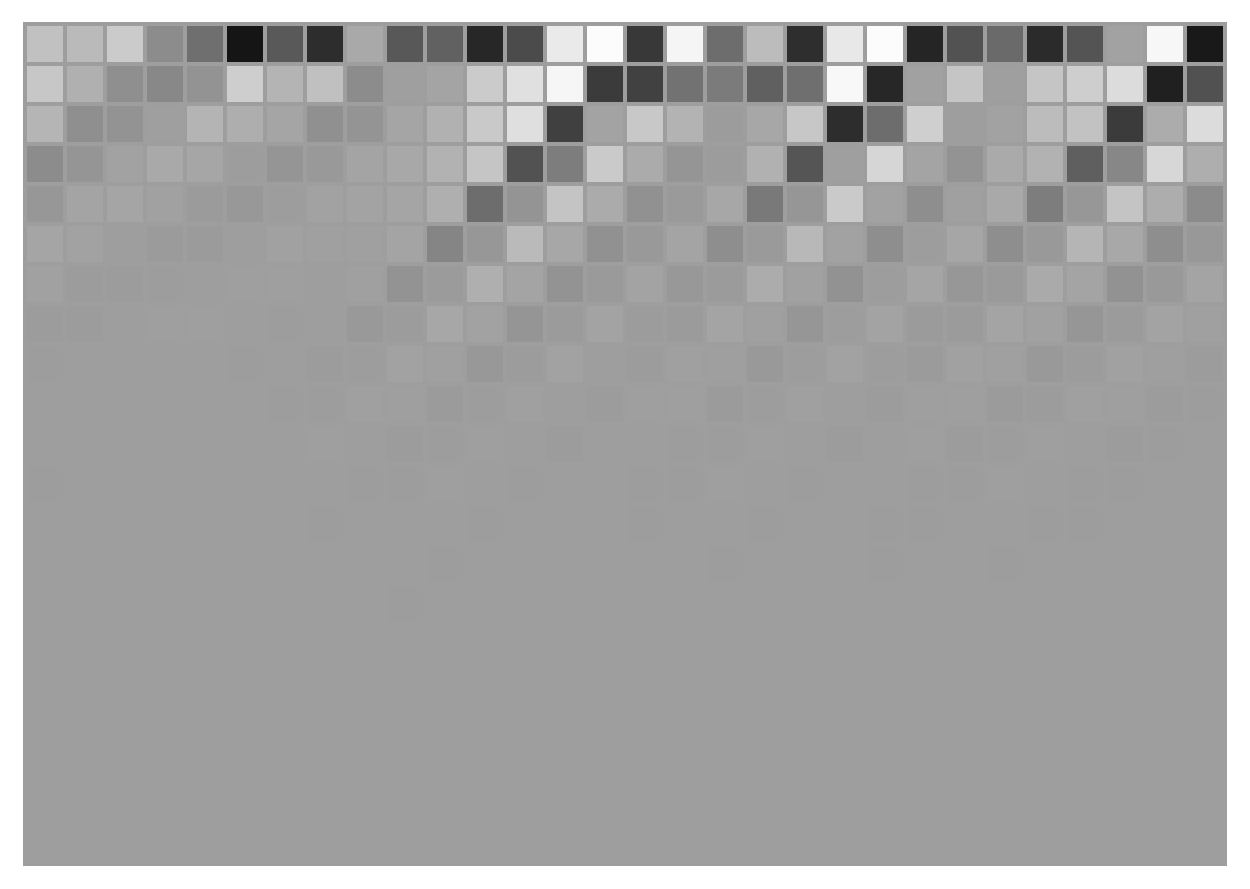}
\end{center}
  \caption{Example of solution to fuzzy ECA defined by (\ref{fuzzy ECA}).  State values are shown in the grayscale from white (0) to black (1).}
  \label{fig:fuzzy ECA38}
\end{figure}

  We can consider (\ref{fuzzy ECA}) is a fuzzy cellular automaton obtained by extending the state value of ECA38 to be continuous in the range $[0,1]$.  Example of evolution from random initial data in $[0,1]$ is shown in Figure~\ref{fig:fuzzy ECA38}.
This figure suggests that the random initial data converges to the uniform state for $n\to\infty$.  If we assume a uniform solution as $u_j^n=v_n$, then $v_n$ satisfies the following mapping,
\begin{equation}
  v_{n+1}=f(v_n,v_n,v_n)=v_n(1-v_n)(2-v_n).
\end{equation}
This mapping can be closed in $[0,1]$ and there is only one stable fixed point $\omega=(3-\sqrt{5})/2$ satisfying $\omega=f(\omega,\omega,\omega)$, that is, $\omega^2-3\omega+1=0$ and $0\le\omega\le1$.  Numerical computations from random initial data as shown in Figure~\ref{fig:fuzzy ECA38} imply that the asymptotic solution converges to the uniform state as $u_j^n\to\omega$.  However, if we restrict the initial data to be binary, that is, 0 or 1, then the uniform state $u\equiv\omega$ can not appear considering the rule table~(\ref{ECA38}) and stable triangular waves with value 1 propagate in $-j$ direction as shown in Figure~\ref{fig:ECA38}.\par
  Asymptotic behaviors of both solutions are very different each other.  One is uniform and static and the other non-uniform and moving stably.  In this article, we discuss and classify the asymptotic solutions to (\ref{fuzzy ECA}) closed in $[0,1]$.  The contents of this article are as follows.  In section~\ref{sec:01}, asymptotic solutions including at least one 0 or 1 are discussed.  We call this type of solution `type A'.  In section~\ref{sec:non01}, those closed in $(0,1)$ including neither 0 nor 1 are discussed.  We call this type of solution `type B'.  In section~\ref{sec:remark}, we give concluding remarks.\par
  There exist related works about the convergence of solutions to fuzzy CA.  Fukuda et.~al. studied asymptotic uniform solutions for some fuzzy CA's\break numerically\cite{fukuda}.  Mingarelli studied a fuzzy CA made from ECA of rule number 110 and showed that a solution from initial data with non-zero value on one site in a zero background converges to a uniform solution\cite{mingarelli2}.
\section{Asymptotic solutions including 0 or 1}  \label{sec:01}
  In this section, we discuss asymptotic solutions of type A, that is, those where there exists at least one site such that $u=0$ or 1 for $n\to\infty$.  Once $0<u_j^n<1$ holds for any $j$ at a certain $n$, the solution always satisfies $0<u<1$ thereafter since (\ref{fuzzy ECA}) can be considered to be an interpolation between $y$ and $1-y$ with weights $(1-x)(1-z)$ and $z$.  Therefore, there exists at least one site with value 0 or 1 at arbitrary $n\gg0$ for asymptotic solutions of type A, and type B otherwise.  Note that a special case of the uniform solution $u\equiv0$ is also of type A but we exclude this trivial case from the discussion below. \par
  Assume the symbol $\st$ denotes an arbitrary value $x$ satisfying $0<x<1$.  Then, the rule table for the different combination of values other than (\ref{ECA38}) is given as follows.
\begin{equation}  \label{fECA38}
\begin{array}{c}
\begin{array}{|c|c|c|c|c|c|c|c|}
\hline
11\st & 10\st & 01\st & 00\st & 1\st1 & 1\st0 & 0\st1 & 0\st0 \\
\hline
0 & \st & \st & \st & \st & 0 &\st & \st \\
\hline
\end{array} \medskip\\
\begin{array}{|c|c|c|c|}
\hline
\st11 & \st10 & \st01 & \st00 \\
\hline
0 & \st & 1 & 0 \\
\hline
\end{array} \medskip\\
\begin{array}{|c|c|c|c|c|c|c|}
\hline
1\st\st & 0\st\st & \st1\st & \st0\st & \st\st1 & \st\st0 & \st\st\st \\
\hline
\st & \st & \st & \st & \st & \st & \st \\
\hline
\end{array}
\end{array}
\end{equation}
Consider all local patterns $\st\st$ with 0 or 1 attached to its left.  Then, the pattern evolves as follows.
\begin{equation}
\begin{array}{lcrrrr}
n\ &:& 0\st\st   \qquad & 01\st\st   \qquad & 11\st\st  \qquad & \st1\st\st \\
n+1&:& \st\st\st \qquad & \st\st\st  \qquad &  0\st\st  \qquad & \st\st\st\st \\
n+2&:&                  &                   & \st\st\st \qquad
\end{array}
\end{equation}
These results mean the following. The sequence of $\st$ grows in the evolution if it includes $\st\st$, the asymptotic solution includes neither 0 nor 1 after enough time steps and $0<u<1$ holds for any $u$'s at $n\gg0$.  Since this type of asymptotic solution is of type B, we will discuss it in the next section.\par
  Thus, if $\st$ is included in the solution of type A, it must be isolated as $0\st0$, $0\st1$, $1\st0$ or $1\st1$.  Consider the local sequence $u_j^n\ldots u_{j+4}^n$ of any combination of 0, 1 and $\st$ which determine the sequence $u_{j+1}^{n+1}u_{j+2}^{n+1}u_{j+3}^{n+1}$ and check what sequence can produce an isolated $\st$.  Using the rule tables (\ref{ECA38}) and (\ref{fECA38}), we can show the following only four patterns can produce it.
\begin{equation}  \label{01*0}
\begin{array}{lccccc}
n &:& 001\st0 \qquad & 101\st0 \qquad & \st01\st0 \qquad & 11\st11 \\
n+1&:& 1\st0   \qquad & 1\st0   \qquad &  1\st0    \qquad & 0\st0 \\
\end{array}
\end{equation}
Therefore, if isolated $\st$'s exist in the asymptotic solution, it must be $01\st0$ moving to $-j$ direction at speed 1.  Note that we use a symbolic calculation program by Mathematica to derive (\ref{01*0}) since the number of cases is large.\par
  Next, we discuss about the sequence of 1's.  Similarly as above, considering the local sequence $u_j^n\ldots u_{j+6}^n$ of any combination of values of 0, 1 and $\st$ and calculating $u_{j+2}^{n+2}u_{j+3}^{n+2}u_{j+4}^{n+2}$ by the program, we can show any combination at $n$ cannot produce 111 at $n+2$.  Therefore, if a sequence of 1's is included in the asymptotic solution, it must be $x11y$ or $x1y$ where $x$ and $y$ are 0 or $\st$.  Moreover, considering the isolated $\st$ is $01\st0$ as shown above, the sequence of 1's must be either of 0110, 0100, 0101 or $01\st0$.  Calculating possible local sequences $u_j^n\ldots u_{j+7}^n$ which produce $u_{j+2}^{n+2}u_{j+3}^{n+2}u_{j+4}^{n+2}u_{j+5}^{n+2}=0110$, 0100 or 0101, we obtain the following evolutions.
\begin{equation}  \label{0110 0100}
\begin{array}{lcll}
n\ &:& \verb/..../0110 & \verb/..../0100\\
n+1&:& \verb/ ../0100  & \verb/ ../0110 \\
n+2&:& \verb/  /0110   & \verb/  /0100
\end{array}
\end{equation}
Note that 0101 cannot be produced and the symbol `\verb/./' denotes an appropriate value with detailed information omitted.\par
\begin{figure}[hbt]
\begin{center}
  \includegraphics[scale=0.5,bb=0 0 360 256]{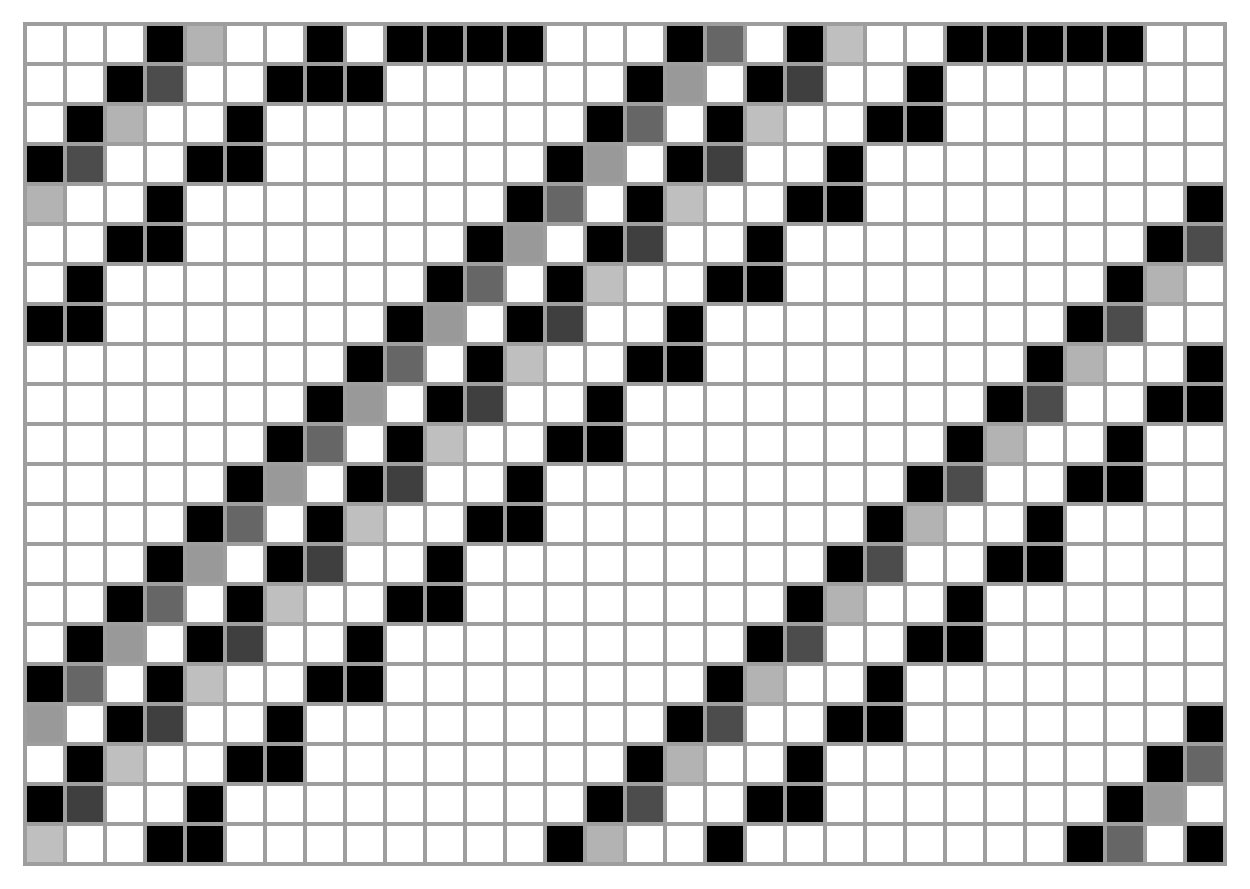}
\end{center}
\caption{Example of time evolution which becomes an asymptotic solution of type A.}  \label{fig:type A}
\end{figure}

  Summarizing the above results, we can conclude about asymptotic solutions of type A as follows.
\begin{itemize}
\item
If $\st$'s exist, they are isolated and realized by the local pattern $01\st0$ moving to $-j$ direction at speed 1.
\item
If 1's exist other than $01\st0$, they are 0110 or 0100.  These two patterns appear alternately as time proceeds and move to $-j$ direction at speed 1.
\item
Among local patterns $1\st$, 11 and 10, one or more 0's exist.
\end{itemize}
An example of time evolution of type A is shown in Figure~\ref{fig:type A}.
\section{Asymptotic solutions with neither 0 nor 1}  \label{sec:non01}
  In this section, we discuss asymptotic solutions of type B, that is, $0<u_j^n<1$ for any $j$ for $n\to\infty$.  We can prove that any $u_j^n$ converges to $\omega=(3-\sqrt{5})/2$ and the solution becomes uniform with a constant $\omega$ as follows.\par
  Assume any pair of $a$ and $b$ satisfying,
\begin{equation}
  0<a\le\omega,\qquad \frac{1-a}{2-a}\le b\le\frac{1-2a}{1-a}.
\end{equation}
Note that $(1-a)/(2-a)\le(1-2a)/(1-a)$ if $0<a\le\omega$, and $a\le b$.  Moreover, the minimum and the maximum of $f(x,y,z)$ in the range $x$, $y$, $z\in[a,b]$ are
\begin{equation}
\begin{aligned}
  & \min_{x,y,z\in[a,b]}f(x,y,z)=f(b,a,a)=a(1-a)(2-b),\\
  & \max_{x,y,z\in[a,b]}f(x,y,z)=f(a,a,b)=(1-a)(a+b-ab).
\end{aligned}
\end{equation}
\par
  Next, let us consider the following sequence for $a_n$ and $b_n$,
\begin{equation}
  a_{n+1}=f(b_n,a_n,a_n),\qquad
  b_{n+1}=f(a_n,a_n,b_n).
\end{equation}
If we assume
\begin{equation}  \label{ab cond}
  0<a_n\le\omega,\qquad \frac{1-a_n}{2-a_n}\le b_n\le\frac{1-2a_n}{1-a_n},
\end{equation}
we can derive $a_n\le a_{n+1}$ and $b_n\ge b_{n+1}$ since we have
\begin{equation}  \label{anbn}
\begin{aligned}
  a_{n+1}-a_n&=a_n(1-a_n)\Bigl(\frac{1-2a_n}{1-a_n}-b_n\Bigr)\ge0, \\
  b_{n+1}-b_n&=a_n(2-a_n)\Bigl(\frac{1-a_n}{2-a_n}-b_n\Bigr)\le0.
\end{aligned}
\end{equation}
Moreover, the same form of inequalities as (\ref{ab cond}) hold for $a_{n+1}$ and $b_{n+1}$ as
\begin{equation}  \label{an+1bn+1}
  0<a_{n+1}\le\omega,\qquad \frac{1-a_{n+1}}{2-a_{n+1}}\le b_{n+1}\le\frac{1-2a_{n+1}}{1-a_{n+1}}.
\end{equation}
Since the proof of (\ref{an+1bn+1}) is not difficult but tedious, we omit it.  Summarizing the above facts, we can conclude the following proposition.
\begin{proposition}  \label{prop1}
  Consider the sequences on $a_n$ and $b_n$ ($n\ge0$)
\begin{equation}
  a_{n+1}=f(b_n,a_n,a_n),\qquad b_{n+1}=f(a_n,a_n,b_n),
\end{equation}
 with initial terms $a_0$ and $b_0$ satisfying
\begin{equation}
 0<a_0<\omega,\qquad \frac{1-a_0}{2-a_0}\le b_0\le\frac{1-2a_0}{1-a_0}.
\end{equation}
Then, $a_n$ and $b_n$ for any $n$ satisfy the same form of inequalities as of $a_0$.  Moreover, $a_n\le a_{n+1}$ and $b_{n+1}\le b_n$ hold and the interval $[a_n,b_n]$ is nested as $[a_{n+1},b_{n+1}]\subset[a_n,b_n]$.
\end{proposition}
\par
  Since the sequence of interval $[a_n,b_n]$ is nested, it converges to $[\alpha,\beta]$ for $n\to\infty$.  Values $\alpha$ and $\beta$ satisfy $\alpha=f(\beta,\alpha,\alpha)$ and $\beta=f(\alpha,\alpha,\beta)$.  The solution of this couple of equations is uniquely determined as $\alpha=\beta=\omega$ in the range of $0<\alpha\le\beta<1$.\par
  Finally, let us consider the asymptotic solution of type B.  Assume a certain time step of the asymptotic solution is $n=0$ without loss of generality.  The solution satisfies $0<u_j^0<1$ for any $j$.  The size of space sites is finite ($0\le j<K$) and the periodic boundary condition is imposed.  Since the space is finite, there exist the maximum $M$ and the minimum $m$ for $\{u_j^n\}_{j=0}^{K-1}$.  Then, we can choose $a_0$ satisfying $0<a_0<\min(m,\omega)$.  The upper bound $\frac{1-2a_0}{1-a_0}$ for $b_0$ of Proposition~\ref{prop1} converges to 1 as $a_0\to0$ and the lower bound $\frac{1-a_0}{2-a_0}$ to $\frac{1}{2}$.  Therefore, we can always choose $a_0$ and $b_0$ satisfying their inequalities of Proposition~\ref{prop1} for any initial data.  Since $a_{n+1}$ and $b_{n+1}$ is the maximum and the minimum of $f(x,y,z)$ in the range of $x$, $y$, $z\in[a_n,b_n]$, $u_j^n\in[a_n,b_n]$ holds for any $n$ from Proposition~\ref{prop1}.  Thus we obtain $\lim_{n\to\infty}u_j^n=\omega$ for any $j$.  It means that the asymptotic solution of type B is a uniform solution with the value $\omega$.
\section{Concluding remarks}  \label{sec:remark}
  We discussed the asymptotic solutions to (\ref{fuzzy ECA}).  The solutions are always classified into two types, type A and B.  We show that the stable propagating wave with local pattern 0110, 0100 or $01\st0$ exists in type A. Binary solutions constructed only from $\{0,1\}$ are classified into this type.  On the other hand, the asymptotic solution of type B is unique as a uniform solution $u\equiv\omega$.\par
  Though the asymptotic solutions are completely classified, it is more difficult to solve the initial value problem of (\ref{fuzzy ECA}) and it is a future problem.  Moreover, the similar results to type B are reported for other fuzzy CA's\cite{fukuda,mingarelli2}.  It is another future problem to analyze these equations and to classify the solutions.\par
  There are various forms of fuzzy CA produced from its original CA.  We have another example of fuzzy CA replacing the term $xyz$ by $xy^2z$ in (\ref{fuzzy ECA}) and confirmed by numerical calculations that the asymptotic solutions are classified into two types similarly to (\ref{fuzzy ECA}).  It is also a future problem to develop a general method to classify fuzzy CA's originated from the same CA according to their behavior of solutions.

\end{document}